\documentclass[%
 reprint,
superscriptaddress,
 amsmath,amssymb,
 aps, physrev,
]{revtex4-2}

\usepackage{graphicx}
\usepackage{subcaption} 
\usepackage{dcolumn}
\usepackage{bm}
\usepackage{mathrsfs}
\usepackage{xcolor}

\begin{document}

\preprint{APS/123-QED}

\title{\textbf{Signals for Nuclear Solid-Liquid Phase Transition in Clustering Nuclei} 
}%

\author{Xi-Guang Cao}
\email{Contact author: caoxg@sari.ac.cn}
\affiliation{Shanghai Advanced Research Institute, Chinese Academy of Sciences, Shanghai 201210, China}
\affiliation{Shanghai Institute of Applied Physics, Chinese Academy of Sciences, Shanghai 201800, China}
\affiliation{University of Chinese Academy of Sciences, Beijing 101408, China.}

\author{Yong-Hao Jin}
\affiliation{Shanghai Institute of Applied Physics, Chinese Academy of Sciences, Shanghai 201800, China}
\affiliation{Shanghai Advanced Research Institute, Chinese Academy of Sciences, Shanghai 201210, China}
\affiliation{University of Chinese Academy of Sciences, Beijing 101408, China.}

\author{Chun-Wang Ma}\email{Contact author: machunwang@126.com}
\affiliation{Institute of Nuclear Science and Technology, Henan Academy of Sciences, Zhengzhou 450046, China}
\affiliation{Centre of Theoretical Physics, College of Physics, Henan Normal University, Xinxiang 453007, China}
\affiliation{Shanghai Research Center for Theoretical Nuclear Physics, NSFC and Fudan University, Shanghai 200438, China}

\author{Wen-Bo Liu}
\affiliation{Shanghai Advanced Research Institute, Chinese Academy of Sciences, Shanghai 201210, China}
\affiliation{Shanghai Institute of Applied Physics, Chinese Academy of Sciences, Shanghai 201800, China}
\affiliation{Centre of Theoretical Physics, College of Physics, Henan Normal University, Xinxiang 453007, China}

\author{Yu-Gang Ma}\email{Contact author: mayugang@fudan.edu.cn}
\affiliation{Key Laboratory of Nuclear Physics and Ion-beam Application (MOE), Institute of Modern Physics, Fudan University, Shanghai, 200433, China}
\affiliation{Shanghai Research Center for Theoretical Nuclear Physics, NSFC and Fudan University, Shanghai 200438, China}
\affiliation{School of Physics, East China Normal University, Shanghai 200062, China}

\begin{abstract}
The nuclear phase diagram remains a fundamental challenge, where uncovering how the cluster degree of freedom evolves with temperature and energy is essential to understand the intriguing crossover between cluster, nucleon, and nuclear astrophysics.
A new first-order phase transition, the nuclear solid-liquid phase transition (SLPT), emerges spontaneously from nucleonic degrees of freedom in low-energy heavy-ion collisions. 
Clear signals of the SLPT are identified as a region of negative heat capacity in smaller systems (e.g., $^{12}\text{C} + ^{12}\text{C}$ and $^{16}\text{O} + ^{12}\text{C}$) and a caloric plateau in larger systems (e.g., $^{28}\text{Si} + ^{12}\text{C}$ and $^{40}\text{Ca} + ^{12}\text{C}$), along with prominent behaviors in the number of fragment species and information entropy.
Moreover, the pairwise distance probability density function provides a direct spatial fingerprint of the melting, where long-range crystalline $\alpha$-cluster peaks vanish above the transition temperature, giving way to a liquid-like distribution.
This new phase transition unifies the many-body dynamics of finite constituents, extending from nanoscale clusters to nuclear systems. 
Beyond advancing nuclear structure and phase diagrams, these insights offer profound implications for astrophysics, notably linking structural evolution to astrophysical phase transitions via nuclear cluster dissolution in supernovae.

\end{abstract}

\maketitle

\section{\label{sec:level1}Introduction}
The thermodynamics of phase transitions in highly finite systems is a fundamentally interdisciplinary frontier, bridging nuclear physics with condensed matter, cold atom physics, and astrophysics~\cite{MoreisDiff, BERRY2013205}.
In atomic and molecular nanoclusters containing as few as several constituents, a rounded, first-order solid-liquid phase transition (SLPT) has been extensively identified, characterized by a size-dependent melting temperature and specific heat anomalies~\cite{PhysRevLett.65.1567, PhysRevLett.77.99, PhysRevLett.94.233401, George_Science_277_5332_1619, PhysRevLett.79.99, schmidt_irregular_1998, briant_molecular_1975}.
By analogy, atomic nuclei provide a distinctly quantum, strongly interacting Fermi-scale laboratory to test the boundaries of these universal scaling laws.
Yet phase transition studies in excited nuclei, governed by the fundamental nature of the nucleon-nucleon interaction, have historically focused intensely on the nuclear liquid-gas phase transition (LGPT) in the high-temperature regime to explore the nuclear equation of state (EoS)~\cite{Poch75PRL, PhysRevLett.83.3617, Bonasera_review_2000, borderie_liquidgas_2019}.
However, it remains a fundamental, open question whether an analogous SLPT can occur in the nuclear low-excitation regime, manifesting as a structural melting from a highly ordered, solid cluster state to a disordered, liquid state.
Crucially, the existence of highly ordered, solid cluster configurations at low excitation energies provides the essential physical environment for such an SLPT to emerge.
It is now well-established that $\alpha$-conjugate nuclei can manifest highly ordered, molecule-like or crystalline spatial symmetries, such as the $\mathcal{D}_{3h}$ arrangement in $^{12}$C~\cite{PhysRevLett.113.012502, Kanada-Enyo:2012yif, Liu_2012}, the tetrahedral configuration in $^{16}$O~\cite{Bijker:2014tka, Epelbaum:2013paa}, and even more complex exotic structures in heavier systems~\cite{Ebran:2012ww}.
Beyond ground-state symmetries, nuclear clustering exhibits a wealth of physical phenomena~\cite{Xu2025PRC, Xu2026PRC}, manifesting exotic configurations that challenge conventional theoretical frameworks, such as bubble and toroidal structures~\cite{wong1972toroidal, WONG1973279, wong1978rotating, staszczak2014region}.
The continuous advancement of high-precision measurement methods is driven by these unresolved fundamental questions, including the exploration and characterization of such exotic cluster configurations~\cite{He:2014iqa, Cao:2018cvf}, the potential existence of $\alpha$-cluster Bose-Einstein condensation at the nucleon level~\cite{PRL.THSR, ZhouB, PRL07}, and the critical influence of cluster structures on the nuclear equation of state (EoS) in low-density astrophysical environments~\cite{PRL12}.
Concurrently, deciphering the underlying exotic structures of $\alpha$-conjugate nuclei has become a focal point of intense interest across both theoretical and experimental fronts over the past decade~\cite{PhysRevC.97.014317, Dey_PRC_2020}.
Reflecting these efforts, notable advances include the deployment of high-resolution charged-particle telescope arrays~\cite{Liao2024NST} and the precise experimental identification of highly symmetric toroidal configurations in $^{28}\text{Si}$ at high excitation energies via multi-alpha decay analyses~\cite{Cao:2018cvf, T.DAldo2025PRC, Natowitz2026PRC}.
These diverse exotic configurations strongly reinforce that $\alpha$-conjugate nuclei can manifest as highly ordered cluster states under specific conditions.

While these comprehensive structural and experimental milestones confirm the robust existence of solid cluster phases, their thermodynamic and dynamical evolution under thermal excitation remains completely unexplored.
Unveiling this novel phenomenon would elucidate how the breaking of cluster symmetries shapes the equation of state and neutrino transport in hot neutron star crusts---processes that profoundly influence crustal transport properties, the dynamics of core-collapse supernovae, and explosive nucleosynthesis.
More fundamentally, the SLPT provides a highly sensitive probe to settle the long-standing controversy over various $\alpha$-cluster configurations and their structural evolution.
By tracking how these cluster degrees of freedom respond to excitation, this transition bridges nuclear structure and low-energy reaction dynamics.
Beyond nuclear physics, exploring such a phase transition in finite-cluster systems advances our interdisciplinary understanding of statistical mechanics and phase dynamics in mesoscopic regimes—spanning cold atoms and nanostructures—where macroscopic limits break down~\cite{MoreisDiff}.

In this work, we report the identification of the SLPT within $\alpha$-conjugate systems.
To accurately track the microscopic structural and phase evolution, we employ the extended quantum molecular dynamics (EQMD) model. 
By simulating the evolution of central collision systems (from $^{12}$C + $^{12}$C to $^{40}$Ca + $^{12}$C), we demonstrate that the melting of $\alpha$-clustering configuration triggers unambiguous thermodynamic and informational anomalies, providing direct signals for the nuclear SLPT.

\section{Methods}
\label{sec:append}

Characterizing a thermodynamic phase transition fundamentally relies on the reliable extraction of state variables, particularly temperature.
However, the extraction of nuclear temperature in intermediate-energy heavy-ion collisions (HICs) remains highly debated—largely due to the difficulty of achieving statistical thermal equilibrium in hot transient nuclei prior to emission.
In contrast, the temperature in low-energy HICs can be defined with substantially reduced ambiguity.
Because the interaction time in low-energy collisions far exceeds the relaxation time of internal nuclear degrees of freedom~\cite{Feldmeier_1987}, the system can fully relax, facilitating the establishment of statistical thermal equilibrium.

To investigate the evolution of cluster degrees of freedom and phase transitions in finite nuclear systems, we perform calculations for central collisions of $^{12}$C + $^{12}$C, $^{16}$O + $^{12}$C, $^{28}$Si + $^{12}$C, and $^{40}$Ca + $^{12}$C at incident energies ranging from 2.0 to 7.0 MeV/nucleon.
The dynamics are described using the extended quantum molecular dynamics (EQMD) model~\cite{Maruyama:1995dc}.
Unlike standard QMD, EQMD incorporates a Pauli potential and treats the wave-packet widths as complex parameters, thereby enabling them to evolve as independent dynamic variables for each nucleon.
By explicitly treating nucleonic degrees of freedom without assuming \textit{a priori} spatial geometries, EQMD has proven highly robust in tracking cluster dynamics via different experimental probes~\cite{He:2014iqa, He:2016cwt, Dey_PRC_2020, REN2024PLB, Huang:2021mct, Wang:2017tfr, HBS95PRC, ZhangS95PRC, MaYG:2022dbh}.

\begin{figure}[htbp]
    \centering
    \begin{subfigure}[t]{0.48\linewidth}
        \centering
        \includegraphics[width=\linewidth]{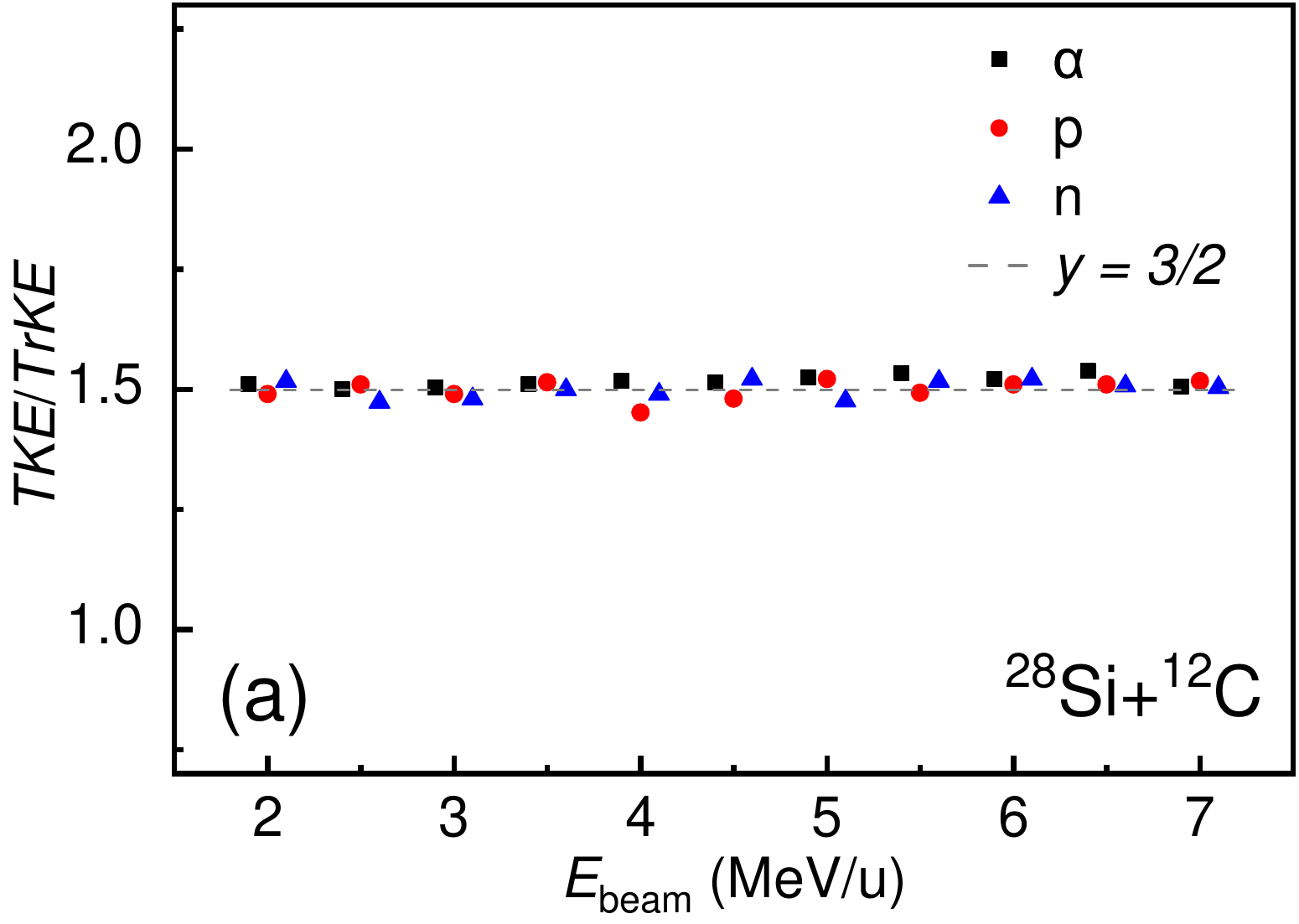}
    \end{subfigure}
    \centering
    \raisebox{0.12cm}{%
        \begin{subfigure}[t]{0.49\linewidth}
            \centering
            \includegraphics[width=\linewidth]{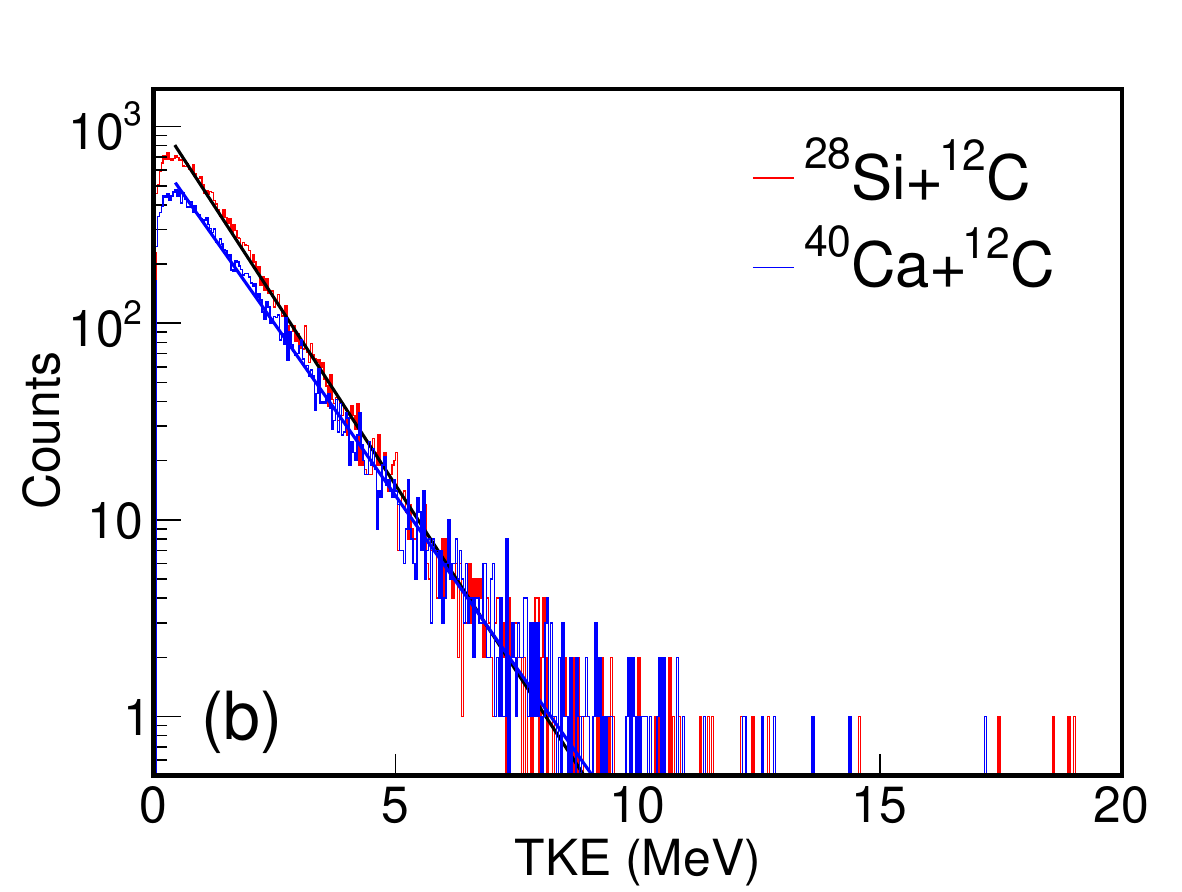}
        \end{subfigure}%
    }
    \captionsetup{justification=raggedright, singlelinecheck=false}
    \caption{(Color online) Kinetic energy characteristics of $\alpha$ clusters, protons, and neutrons. (a) The ratio of total kinetic energy (TKE) to transverse kinetic energy (TrKE) of $\alpha$ clusters, protons, and neutrons produced in central collisions of $^{28}$Si + $^{12}$C. The horizontal line at $3/2$ is drawn to guide the eye. (b) The TKE spectra of $\alpha$ clusters and fits using a Maxwell-Boltzmann distribution.}
    
    \label{fig:equilibrum and KE}
\end{figure}

Crucially for the present study, the advanced treatment within EQMD accurately reproduces nuclear ground-state properties and ensures the remarkable long-term stability of initialized nuclei~\cite{wangssNPR_2015}. 
Thanks to this stability, spurious nucleon emissions are effectively prevented, allowing the compound system to reach statistical thermal equilibrium at around several thousand fm/c after projectile-target contact. 
This equilibration is firmly validated by two key criteria.
First, the internal momentum distribution of the compound system is highly isotropic; this is evidenced by the fact that the ratio of the total kinetic energy (TKE) to the transverse kinetic energy (TrKE) is approximately 3/2 for the constituent neutrons, protons, and $\alpha$ clusters, as shown in Fig.~\ref{fig:equilibrum and KE}(a).
Second, the total kinetic energies of these particles exhibit an excellent Maxwell-Boltzmann distribution. As illustrated in Fig.~\ref{fig:equilibrum and KE}(b) for $\alpha$ clusters produced in the central collisions of $^{28}$Si + $^{12}$C and $^{40}$Ca + $^{12}$C, these equilibrated spectra serve as a robust slope thermometer.
By fitting the $\alpha$-cluster TKE distributions using Eq.~(\ref{eq.YE}), we reliably extract the thermodynamic temperature $T$ of the transient hot system.

\begin{equation}
    Y(E)\propto \exp\left(-\frac{E}{T}\right).
    \label{eq.YE}
\end{equation}

To construct the caloric curve, the extracted temperature must be plotted against the corresponding energy variable.
While the total excitation energy $E^*$ is conventionally used, it often encompasses non-thermalized dynamical components, primarily collective motion. 
Removing these non-thermal contributions yields the pure thermal energy $E_{\text{th}}$, which more rigorously reflects the true thermodynamic state of the system. 
Therefore, we characterize the system using $E_{\text{th}}$ rather than total excitation energy to isolate purely thermal effects~\cite{Zheng:2010kg}:

\begin{equation}
    \left\langle\frac{E_{\text{th}}}{A}\right\rangle = \frac{E_{\text{c.m.}}}{A} - \left[ \left\langle\frac{E}{\bar{N}}\right\rangle - \frac{3}{2}\left\langle\frac{E_{xy}}{\bar{N}}\right\rangle \right] - Q,
    \label{eq.Eth}
\end{equation}
where $\bar{N}$ is the mean multiplicity, and $Q$ is the reaction $Q$-value. The specific heat is then derived as 
\begin{equation}
    C=\mathrm{d}E_{\text{th}}/\mathrm{d}T.
    \label{Eq.Caloric}
\end{equation}

\section{Results and Discussion}

\begin{figure}[htbp]
    \includegraphics[width=1.0\hsize]{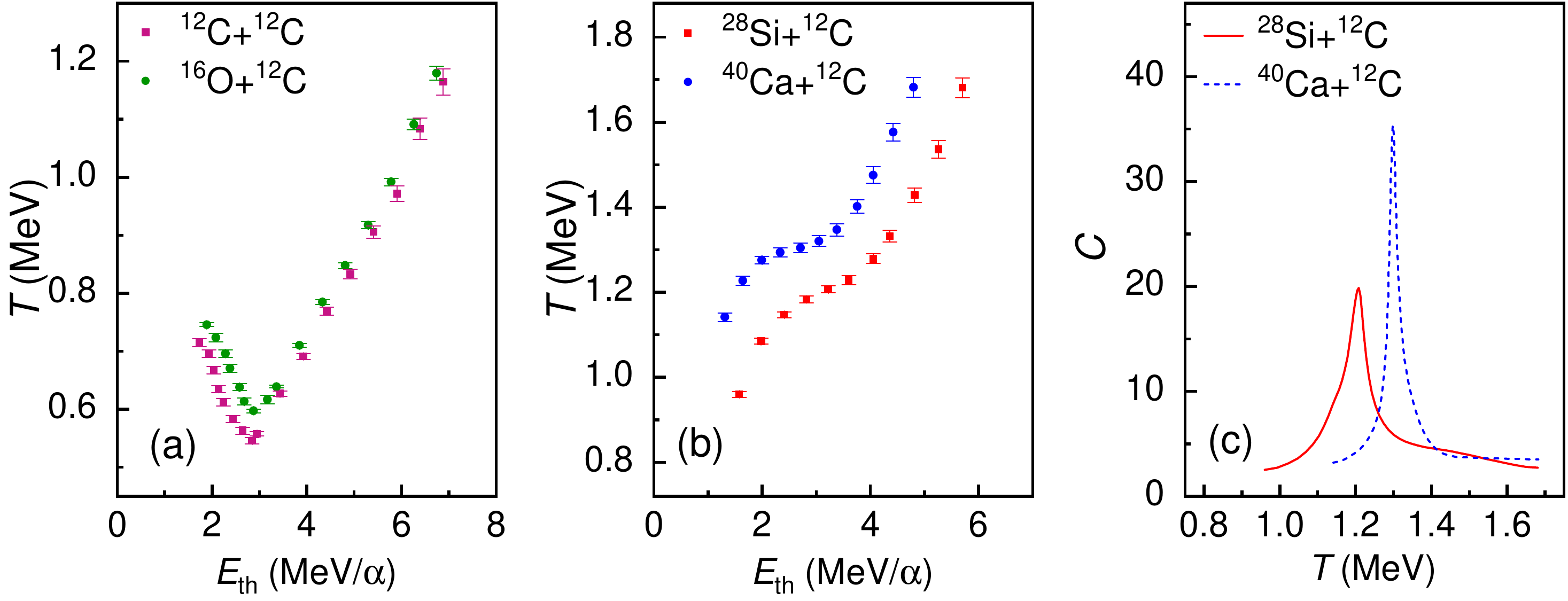}
    \captionsetup{justification=raggedright, singlelinecheck=false}
    \caption{(Color online) Caloric curves and specific heat. (a) Temperature versus thermal energy ($E_{\text{th}}$) per $\alpha$ cluster [Eq.~(\ref{eq.Eth})] for central collisions of $^{12}$C + $^{12}$C and $^{16}$O + $^{12}$C. (b) Corresponding caloric curves for $^{28}$Si + $^{12}$C and $^{40}$Ca + $^{12}$C. (c) Specific heat capacity ($C$) for the heavier systems shown in (b) extracted using Eq.~(\ref{Eq.Caloric}).}
    \label{fig:cal_hea}
\end{figure}

By plotting the extracted temperature against this pure thermal energy across a wide range of incident energies, a clear thermodynamic picture of the phase transition emerges.
The thermodynamic signatures of the $\alpha$-cluster SLPT are presented in Fig.~\ref{fig:cal_hea}.
Focusing first on the thermal response, the caloric curves (Fig.~\ref{fig:cal_hea}(a)(b)) exhibit a striking system-size dependence.
For lighter systems ($^{12}$C + $^{12}$C and $^{16}$O + $^{12}$C), a pronounced "back-bending" structure emerges.
This negative specific heat is a distinct hallmark of a microcanonical first-order phase transition.
As structural melting opens up a vast structural phase space, the isolated system redistributes its internal energy, consuming kinetic energy to break $\alpha$-$\alpha$ bonds and create new fragment surfaces. 
This requisite energy repartitioning inevitably lowers the nuclear temperature even as the total thermal energy increases.

As the system size increases ($^{28}$Si + $^{12}$C and $^{40}$Ca + $^{12}$C), a significant evolution is observed in the caloric curves.  
The pronounced "back-bending" characteristic of the light systems gradually transitions into a distinct plateau structure in heavier systems, as shown by comparing Figs.~\ref{fig:cal_hea}(a) and \ref{fig:cal_hea}(b).  
This evolution is fundamentally driven by the decreasing surface-to-volume ratio of the finite systems~\cite{GROSS1997119}.
As the nucleus becomes heavier, the relative energetic penalty for creating new fragment surfaces is mitigated. 
Consequently, with extreme finite-size effects diminishing, the region of negative specific heat gives way to a broad phase-coexistence plateau—a classic manifestation of a first-order phase transition in bulk-like systems.
Since the temperature remains nearly constant across this plateau ($dT/dE_{\text{th}} \approx 0$), the specific heat $C$ derived from Eq.~(\ref{Eq.Caloric}) correspondingly exhibits a sharp peak [see Fig.~\ref{fig:cal_hea}(c)], signaling a rapid transition in the $\alpha$-cluster state in these larger systems.

From these caloric curves, the extracted transition temperatures are $T_c \approx 0.56$, 0.60, 1.20, and 1.29 MeV for $^{12}$C+$^{12}$C, $^{16}$O+$^{12}$C, $^{28}$Si+$^{12}$C, and $^{40}$Ca+$^{12}$C, respectively, manifesting a systematic scaling with system size.
Notably, this systematic thermodynamic evolution aligns fundamentally with the scaling laws of finite-system phase transitions. 
The profound influence of system size on caloric curves and melting behaviors has been rigorously established across various finite-size systems, spanning foundational theoretical models of nanoscale clusters~\cite{PhysRevLett.65.1567}, experimental calorimetric measurements of sodium clusters~\cite{PhysRevLett.79.99, schmidt_irregular_1998}, and molecular dynamics simulations of Lennard-Jones droplets~\cite{briant_molecular_1975}.

Beyond the phase transition region, the caloric curves for all studied systems exhibit a linear and monotonically increasing behavior. 
This higher-energy linear regime corresponds to a liquid state of the nuclear system.  
At this stage, intense thermal motions dominate the system, and the nuclear potential energy approaches saturation. 
Any additional excitation energy is then primarily converted into the thermal kinetic energy of the constituents.   
This energetic mechanism restores the microcanonical heat capacity to a stable positive value, which manifests as a constant positive slope in the post-transition caloric curves.

To understand the microscopic mechanisms driving these phase behaviors, we examine the dynamical evolution of individual $\alpha$-clusters.
At higher incident energies, the inter-cluster degrees of freedom are excited, dominating the energy deposition within the reaction system.
This excitation induces intense relative motion and structural fluctuations among the $\alpha$ clusters, leading to the breaking of $\alpha$-$\alpha$ bonds even while the average inter-cluster distance remains relatively stable.
Our analysis reveals that the average number of bonds per cluster in these collision systems remains significantly lower than that in their respective ground-state configurations.  
This marked reduction in the average bond number across all studied systems provides direct microscopic evidence for a substantial structural reorganization, signaling the structural melting of the localized $\alpha$-cluster geometry into a spatially delocalized configuration.

Consequently, this profound structural melting and its associated configurational disordering directly reshape the statistical distribution of fragment multiplicities.
This statistical reorganization allows the SLPT to be naturally manifested through informational observables, alongside standard thermodynamic ones.  
Just as multiplicity anomalies serve as a typical signature for the nuclear LGPT~\cite{Chaudhuri_PRC_2019}, their analogous statistical manifestations provide a powerful and sensitive probe for this structural melting in $\alpha$-cluster systems.
To establish a comprehensive multi-probe picture of the phase transition, we extend the informational framework from the LGPT to the present solid-liquid scenario by introducing the information entropy~\cite{Shannon_1948, PhysRevLett.83.3617}, defined as
\begin{equation}
    H = -\sum_{k=1}^{n} p_k \ln p_k,
    \label{eq.shannon}
\end{equation}
where $p_k$ represents the normalized yield probability of the $k$-th distinct fragment species, defined as $p_k = N_k / N_{\text{tot}}$. Here, $N_k$ and $N_{\text{tot}}$ denote the multiplicity of species $k$ and the total fragment multiplicity, respectively. The values of $H$ presented in the subsequent results are statistically averaged over the generated event ensemble at each temperature.

Before presenting the results, we clarify the distinction between thermodynamic entropy and information entropy.
While both thermodynamic and information entropies increase monotonically with temperature, they probe fundamentally distinct degrees of freedom. 
Thermodynamic entropy is heavily dominated by the continuous expansion of momentum space, whereas the information entropy $H$ filters out these thermal backgrounds to strictly isolate the discrete configurational evolution of fragment species.
Near the phase transition, enhanced cluster dissociation and intense structural instabilities lead to a rapid broadening of the fragment species distribution, manifesting as a sharp acceleration in $H$ and consequently producing a pronounced peak in $\mathrm{d}H/\mathrm{d}T$.

\begin{figure}[htbp]
    \includegraphics[width=1.0\columnwidth]{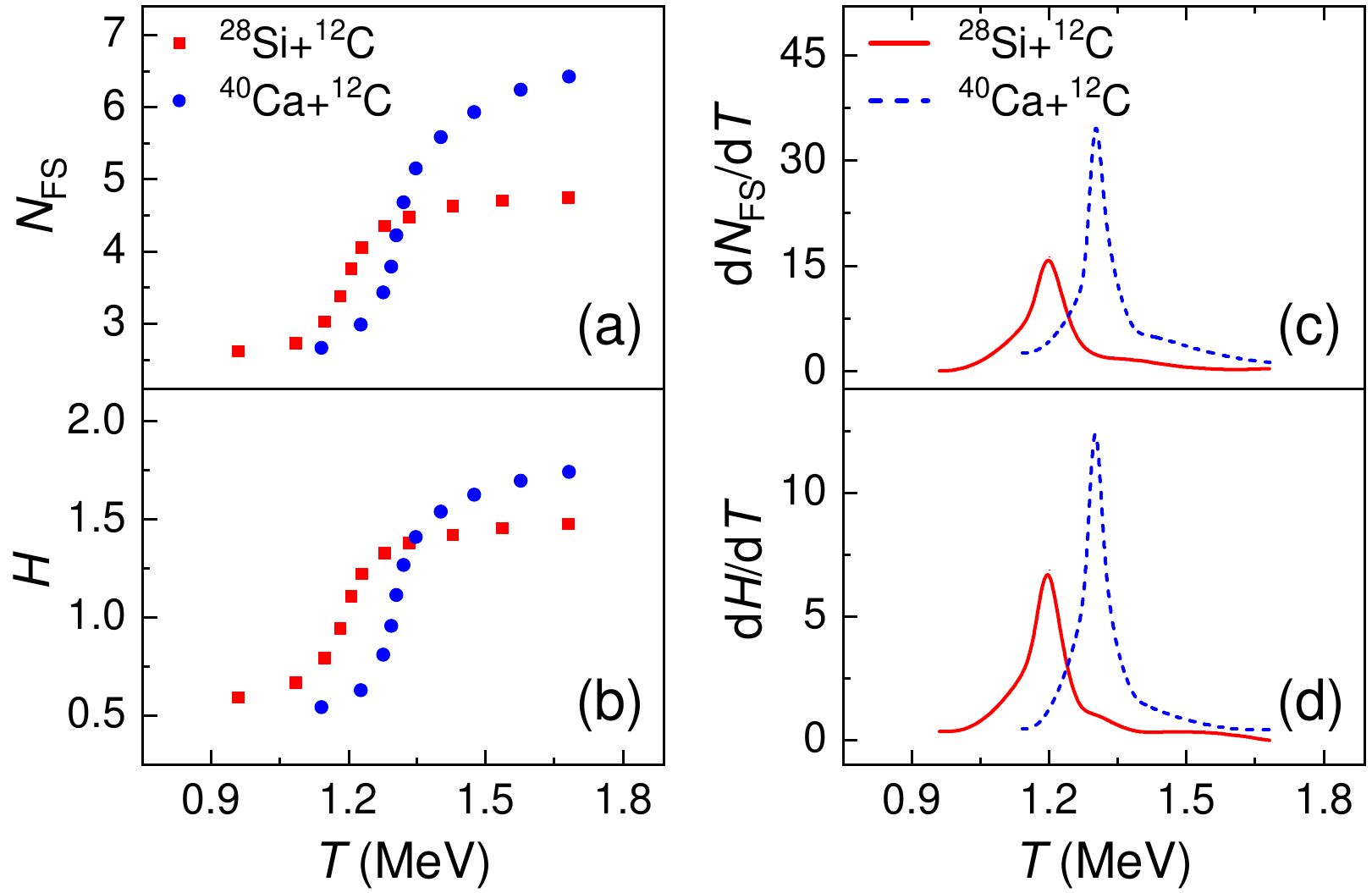}
    \captionsetup{justification=raggedright, singlelinecheck=false}
    \caption{(Color online) Temperature dependence of (a) the number of fragment species $N_{\text{FS}}$ and (b) the information entropy $H$ for the central collisions of $^{28}$Si + $^{12}$C and $^{40}$Ca + $^{12}$C. The corresponding temperature derivatives, (c) $\mathrm{d}N_{\text{FS}}/\mathrm{d}T$ and (d) $\mathrm{d}H/\mathrm{d}T$, are also shown.}
    \label{fig:M_entropy}
\end{figure}

Specifically, the temperature-dependent profiles of the total number of fragment species ($N_{\text{FS}}$) and the information entropy ($H$)—an informational metric previously established as a robust indicator for the LGPT~\cite{PhysRevLett.83.3617}—are evaluated for the simulated central collisions of the heavier systems $^{28}$Si + $^{12}$C and $^{40}$Ca + $^{12}$C, as illustrated in Figs.~\ref{fig:M_entropy}(a) and \ref{fig:M_entropy}(b). 
Both $N_{\text{FS}}$ and $H$ exhibit a strikingly synchronized three-stage evolution, highlighted by a region of rapid, highly nonlinear acceleration within a narrow temperature window.
Microscopically, at low temperatures corresponding to the solid state, the $\alpha$-cluster structure remains largely intact and ordered, resulting in a constrained configuration space characterized by a restricted set of distinct cluster species.
However, once the thermal energy overcomes the inter-cluster binding threshold, this constraint is lifted, reflecting a rapid statistical randomization and the chaotic expansion of the available phase space.
The hot transient nucleus abruptly disintegrates into an abundance of smaller, lighter cluster fragments. 
This sudden and explosive expansion of the available configuration space drives the dramatic, concurrent leap observed in both fragment diversity and Shannon information entropy.

Crucially, the temperature derivatives of these observables, $\mathrm{d}N_{\text{FS}}/\mathrm{d}T$ and $\mathrm{d}H/\mathrm{d}T$, map out this structural transition with high fidelity, displaying sharp, prominent peaks (Figs.~\ref{fig:M_entropy}(c) and \ref{fig:M_entropy}(d)).  
Echoing the behavior observed in the thermodynamic specific heat, the sharpness of these derivative peaks exhibits a clear system-size dependence, becoming more pronounced for the heavier $^{40}$Ca + $^{12}$C system.
Most notably, the transition temperatures extracted from these informational extrema—$T_c \approx 1.20$ MeV and $1.30$ MeV for $^{28}$Si + $^{12}$C and $^{40}$Ca + $^{12}$C, respectively—are in excellent quantitative agreement with the peaks of specific heat ($C$) derived from the caloric curves (Fig.~\ref{fig:cal_hea}(c)). 
The precise alignment between the specific heat peak and the maximum of the temperature derivative of information entropy ($\mathrm{d}H/\mathrm{d}T$) provides robust cross-validation, significantly reinforcing the reliability of the identified phase transition.

\begin{figure}[htbp]
    \includegraphics[width=1.0\columnwidth]{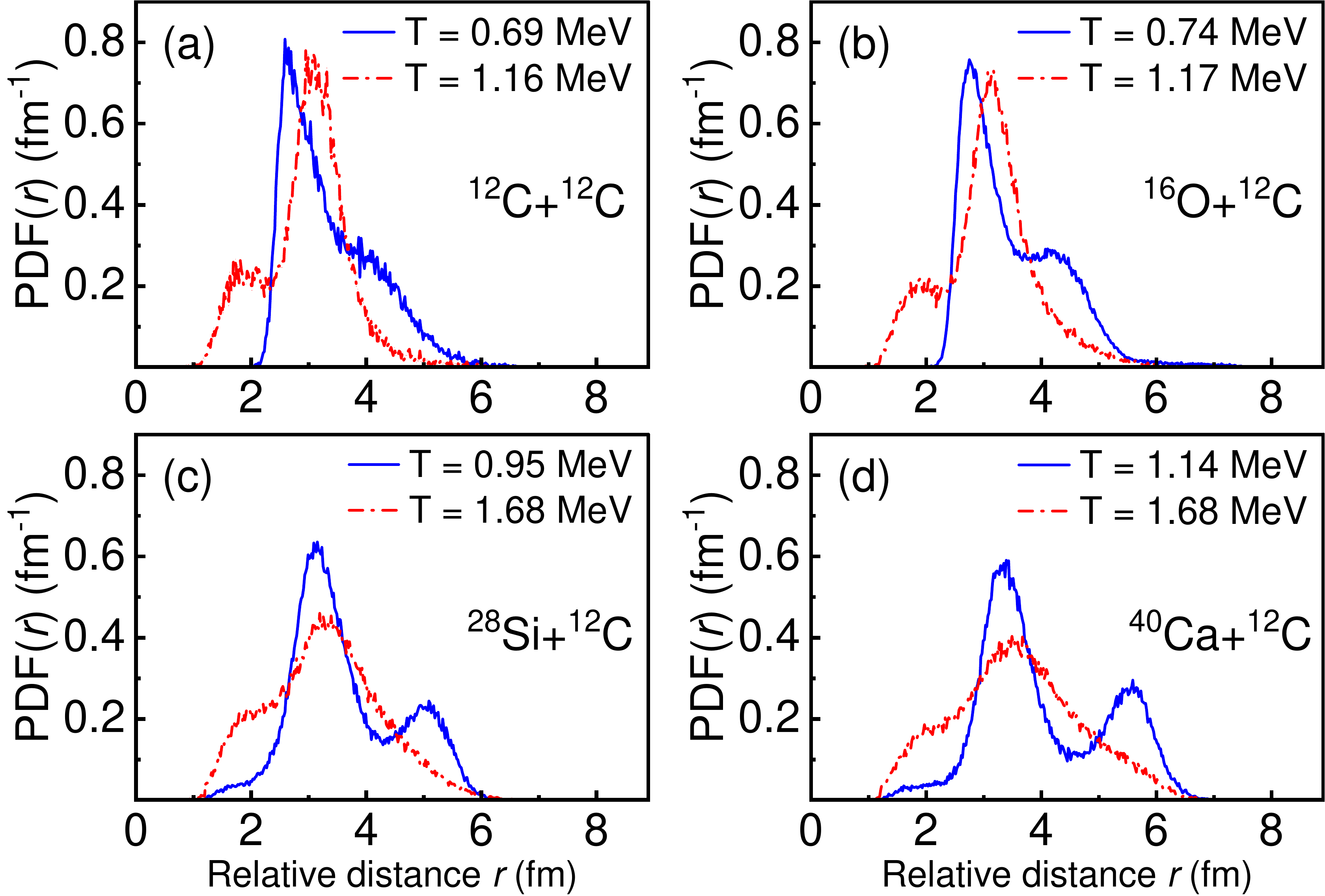}
    \caption{(Color online) Probability density function $\text{PDF}(r)$ of the relative distance between all identified clusters for four collision systems. Solid and dashed curves denote low-temperature ($T < T_c$) and high-temperature ($T > T_c$) states, respectively. } 
    \label{fig:pdf_distance}
\end{figure}
The thermodynamic and informational signatures discussed above establish the existence of the SLPT and its transition temperature.
We now examine the direct spatial manifestation of this transition—how the internal configuration of the system is reorganized across $T_c$.
The pairwise distance probability density function $\text{PDF}(r)$ between all constituents of the system---including $\alpha$-clusters, nucleons, and other clusters---provides a direct picture of the spatial evolution of the excited states.
As shown in Fig.~\ref{fig:pdf_distance}, $\text{PDF}(r)$ evolves from a prominently peaked profile at low temperature, reflecting ordered $\alpha$-cluster solid states, to a less-structured distribution governed by the finite-size geometric constraint and local inter-nucleon spacing above $T_c$.
Specifically, as the system melts, the long-range crystalline peaks thoroughly vanish, giving way to a pronounced short-range inter-nucleon peak and a broad, featureless geometric distribution shaped by disordered nucleons.
This structural evolution from discrete, crystalline correlation peaks to a continuous, liquid distribution constitutes the direct spatial fingerprint of the cluster melting picture.

From an experimental standpoint, directly extracting microcanonical temperatures, caloric curves, and exact specific heat capacities remains a highly challenging endeavor due to detector acceptance limitations and secondary decay corrections.  
Conversely, both fragment diversity ($N_{\text{FS}}$) and Shannon information entropy ($H$) are straightforwardly reconstructible from measured fragment yield distributions, with $H$ inherently quantifying the average configurational disorder and structural complexity of the fragmented system. 
The robust correspondence between these informational derivative peaks and the thermodynamic transition points demonstrates that monitoring fragment species diversity and entropy fluctuations using existing high-resolution charged-particle multi-detector arrays can provide a highly feasible and direct experimental strategy for identifying SLPT signals in $\alpha$-conjugate nuclei.

\section{Summary and Conclusion}

In summary, we have uncovered a novel SLPT in reaction systems induced by $\alpha$-conjugate nuclei ($^{12}$C, $^{16}$O, $^{28}$Si and $^{40}$Ca).
Based on nucleonic degrees of freedom, we establish a multi-observable signature to unambiguously characterize this phase transition.
On the thermodynamic side, the finite-size evolution of the caloric curves reveals a clear transition from the pronounced negative specific heat (``back-bending'') in light systems to an extended plateau in heavier ones.
This thermodynamic evolution is intrinsically coupled to the $\alpha$-cluster structure.  
As the excitation energy increases, the inter-cluster degrees of freedom are activated, leading to the disruption of the spatial symmetry established by the cluster structure and a subsequent transition from an ordered solid cluster state to a disordered liquid phase.

Complementing this calorimetric evidence, the information entropy ($H$) and the total number of fragment species ($N_{\text{FS}}$) provide a powerful structural confirmation of the transition as robust, experimentally accessible observables.
The temperature derivatives, $\mathrm{d}H/\mathrm{d}T$ and $\mathrm{d}N_{\text{FS}}/\mathrm{d}T$, exhibit sharp maxima that precisely coincide with the specific heat ($C$) peaks.
This remarkable convergence among independent thermodynamic and informational probes confirms that the most rapid growth of configurational disorder and structural complexity is a definitive signature of the SLPT. 
In parallel, the pairwise distance distribution directly captures the melting, as sharp long-range crystalline $\alpha$-cluster peaks broaden into a liquid-like profile above $T_c$.
Furthermore, the systematic scaling of the transition temperature ($T_c$) with system size effectively manifests finite-size effects at the nucleonic scale, highlighting the universality of phase transition scaling laws across vastly different scales of matter.

Ultimately, this work reveals that the SLPT of $\alpha$ clusters occurs at distinctly lower temperatures than the conventional nuclear LGPT, while concurrently providing a feasible experimental strategy to identify these transitions using existing multi-detector arrays in low-energy heavy-ion collisions.
This discovery may shed new light on the dynamical and thermodynamic behaviors of $\alpha$-cluster nuclei in extreme astrophysical environments, as well as their configurations, symmetries, and states.
Beyond nuclear systems, our findings offer valuable insights for understanding strongly correlated finite many-body problems in atomic, molecular, and condensed matter physics, where analogous cluster degrees of freedom govern collective phase behaviors.

\begin{acknowledgments}
This work was supported by the National Key Research and Development Program of China (Grant No.
2022YFA1602404), the Strategic Priority Research Program of the Chinese Academy of Sciences (Grant No.
XDB34030000), the National Natural Science Foundation of
China (Grants No. 12475134, No. 12147101, No. 11975210,
and No. U1832129), and the Youth Innovation Promotion
Association CAS (Grant No. 2017309).
\end{acknowledgments}

\appendix

\bibliography{RF_new}

\end{document}